\documentclass{PoS}

\newcommand{\preprint}{
  \begin{picture}(0,0)
    \put(400,170){{\rm\normalsize ADP-09-14/T692}}
  \end{picture}}

\title{\preprint%
Impact of stout-link smearing in lattice fermion actions}

\ShortTitle{Impact of Stout-link smearing in lattice fermion actions}

\author{Peter J. Moran$^a$, Patrick O. Bowman$^b$, \speaker{Derek
    B. Leinweber}$^a$, Anthony G. Williams$^a$ and J. B. Zhang$^{ac}$\\
\llap{$^a$}Special Research Center for the Subatomic Structure of
   Matter (CSSM) and\\Department of Physics, University of Adelaide
   5005, Australia\\
\llap{$^b$}Centre for Theoretical Chemistry and Physics and
  Institute of Natural Sciences,\\
  Massey University (Albany), Private Bag 102904,\\
  North Shore City 0745, New Zealand\\
\llap{$^c$}ZIMP and Department of Physics, Zhejiang University,
   Hangzhou, 310027,\\People's Republic of China\\
E-mail: \email{peter.moran@adelaide.edu.au},
\email{P.O.Bowman@massey.ac.nz},
\email{dleinweb@physics.adelaide.edu.au},
\email{anthony.williams@arcs.org.au},
\email{jzhang@physics.adelaide.edu.au}}

\abstract{The impact of stout-link smearing in lattice fermion actions
  is examined through the consideration of the mass and
  renormalization functions of the overlap quark propagator over a
  variety of smeared configurations.  Up to six sweeps of stout-link
  smearing are investigated.  For heavy quark masses, the quark
  propagator is strongly affected by the smearing procedure.  For
  moderate masses, the effect appears to be negligible. A small effect
  is seen for light quark masses, where dynamical mass generation is
  suppressed through the smearing procedure.}

\FullConference{The XXVII International Symposium on Lattice Field
  Theory - LAT2009\\ July 26-31 2009\\ Peking University, Beijing,
  China}

\begin{document}

\section{Introduction}

The term ``fat-link fermions'' is used to describe a class of actions
in which smeared, or fat, gauge links are incorporated into the
standard Wilson or Clover actions.  Because the smearing algorithms
remove short-distance physics from the gauge field, these actions are
also often referred to as UV-filtered actions.  There is some
flexibility in how these fat links are included in the Dirac operator.
They can be used in all terms of the
action~\cite{DeGrand:1998jq,DeGrand:1998mn,Hasenfratz:2007rf,Capitani:2006ni,Durr:2008rw},
in just the relevant terms~\cite{Cundy:2009yy}, or only in the
irrelevant
terms~\cite{Zanotti:2001yb,Zanotti:2004dr,Boinepalli:2004fz,Kamleh:2007bd}.
The fat-link irrelevant clover (FLIC) action is an example of the last
of these choices.

The FLIC action is
$\mathcal{O}(a)$-improved~\cite{Zanotti:2004dr,Kamleh:2007bd}, and by
only including fat-links in the irrelevant Wilson and Clover terms,
the short-distance physics in the relevant operators is preserved.
This allows relatively cheap access to the light quark mass
regime~\cite{Boinepalli:2004fz}, and excellent scaling with small
$\mathcal{O}(a^2)$ errors that provide near continuum results at a
finite lattice spacing $a$~\cite{Kamleh:2007bd}.

S. D\"urr \emph{et al.} advocate~\cite{Durr:2008zz} a fat-link fermion
action in which all gauge links are smeared.  Using a tree level,
clover improved Wilson action with six sweeps of stout-link smearing,
they successfully reproduced the light hadron spectrum.  While six
sweeps of smearing may seem excessive, there is no clear prescription
for determining the correct number of smearing sweeps.  Indeed,
smearing only introduces extra irrelevant terms into the action, and
so as long as the number of sweeps and the smearing parameter are held
fixed, the smearing should not alter the continuum limit of the
action.

Although not a direct physical observable, the quark propagator is a
fundamental component of QCD, with many physical quantities depending
on its properties.  By studying the momentum-dependent quark mass
function in the infrared region, one can gain valuable insights into
the mechanisms of dynamical chiral symmetry breaking and the
associated dynamical generation of mass.

In the following, we investigate the properties of the momentum space
overlap quark propagator over a variety of smeared quenched
configurations.  Of particular interest is how the mass and
renormalization functions of the fermion propagator change as the
gauge fields undergo more smearing.

As the configurations are smeared, the short-distance, ultraviolet
fluctuations are removed from the gauge field.  For heavy quarks one
might expect that the Compton wavelength would be short enough to
reveal the void in gluon interactions left by the smearing procedure.
As more smearing sweeps are applied, this should cause the asymptotic
value of the mass function to approach the bare input mass.

The ultraviolet physics that is most affected by the smearing
algorithm is linked to the infrared physics through approximate zero
modes.  In the QCD vacuum, there is a finite density of zero modes of
the Dirac operator, $\rho(0)$, due to the presence of topologically
nontrivial, instanton-like objects.  The Dirac zero modes are
intimately related to dynamical chiral symmetry breaking via the
Banks-Casher relation,
\begin{equation}
  \langle \bar{q}q \rangle = -\pi\,\rho(0),\ \mathrm{as}\ m_q
  \rightarrow 0 \,.
\end{equation}
One might therefore expect that at light quark masses, the removal of
short-distance, nontrivial topological fluctuations through smearing
would suppress the generation of approximate zeromodes.  This would
compromise dynamical mass generation, through dynamical chiral
symmetry breaking, for fat-link fermion actions.

\section{Overlap quark propagator}

The massive overlap operator can be written as~\cite{Edwards:1998wx}
\begin{eqnarray}
D(\mu) = \frac{1}{2}\left[1+\mu+(1-\mu)\gamma_5 \epsilon(H_w)
\right] \, , \label{D_mu_eqn}
\end{eqnarray}
where $H_w(x,y)=\gamma_5 D_w(x,y)$ is the Hermitian Wilson-Dirac
operator,  $\epsilon(H_w)$ = $H_w/\sqrt{H_w^2}$ is the matrix sign
function, and the dimensionless quark mass parameter $\mu$ is
\begin{equation}
\mu \equiv \frac{m^0}{2m_w} \, , \label{mu_defn}
\end{equation}
where $m^0$ is the bare quark mass and $m_w$ is the Wilson quark
mass which, in the free case, must lie in the range $0 < m_w < 2$.

In a covariant gauge, the continuum, renormalized quark propagator has
the form
\begin{equation}
  S_\zeta(p) = \frac{Z_\zeta(p^2)}{i{p\!\!\!\!\slash}+M(p^2)} \,.
\end{equation}
Whereas on the lattice we have at tree level,
\begin{equation}
  S^{(0)}(p) \equiv \frac{1}{i{q\!\!\!\!\slash}+m^0} \,,
\end{equation}
where we have defined the kinematic lattice momentum $q$.  The lattice
mass and renormalization functions are then defined
by~\cite{Zhang:2009jf},
\begin{equation}
  S_\zeta(p) = \frac{Z_\zeta(p^2)}{i{q\!\!\!\!\slash}+M(p^2)} \,.
\end{equation}

\section{Simulation details}

We consider $16^3 \times 32$ lattices, generated with a
tadpole-improved L\"uscher-Weisz action, with a lattice spacing of $a
= 0.093$~fm~\cite{Bonnet:1999gt}.  To these configurations we apply
standard, isotropic stout-link smearing~\cite{Morningstar:2003gk} with
$\rho = 0.1$, to create ensembles of 1, 3, and 6-sweep smeared
configurations.  Each level of smearing is then fixed to
$\mathcal{O}(a^2)$-improved Landau gauge~\cite{Bonnet:1999mj}.

We use the mean-field improved Wilson action in the overlap fermion
kernel. The value $\kappa$ = $0.19163$ is used in the Wilson action,
which provides $m_w a$ = $1.391$ for the Wilson regulator mass in the
interacting case~\cite{Bonnet:2002ih}. The overlap quark propagator is
calculated for $15$ bare quark masses on each ensemble and we report
results for three different bare quark masses; $m_0 = 53$~MeV (light),
$177$~MeV (moderate), and $531$~MeV (heavy).  All data is cylinder
cut~\cite{Leinweber:1998im}. Statistical uncertainties are estimated via a
second-order, single-elimination jackknife.

In a standard lattice simulation, one begins by tuning the value of
the input bare quark mass $m^0$ to give the desired renormalized quark
mass.  However, smearing a lattice configuration filters out the
ultraviolet physics and gives a different renormalized quark mass.
The input bare quark mass must then be re-tuned in order to reproduce
the same physical behavior as on the unsmeared configuration.  We
replicate this re-tuning procedure in our results by interpolating
between neighboring quark masses, such that the functions agree at a
given reference momentum $\zeta$.  For full details see
Ref.~\cite{Zhang:2009jf}.

\section{Results}

\begin{figure}
  \begin{center}
    \begin{tabular}{c@{\hspace{0.05\textwidth}}c}
      \includegraphics[angle=90,width=0.36\textwidth]{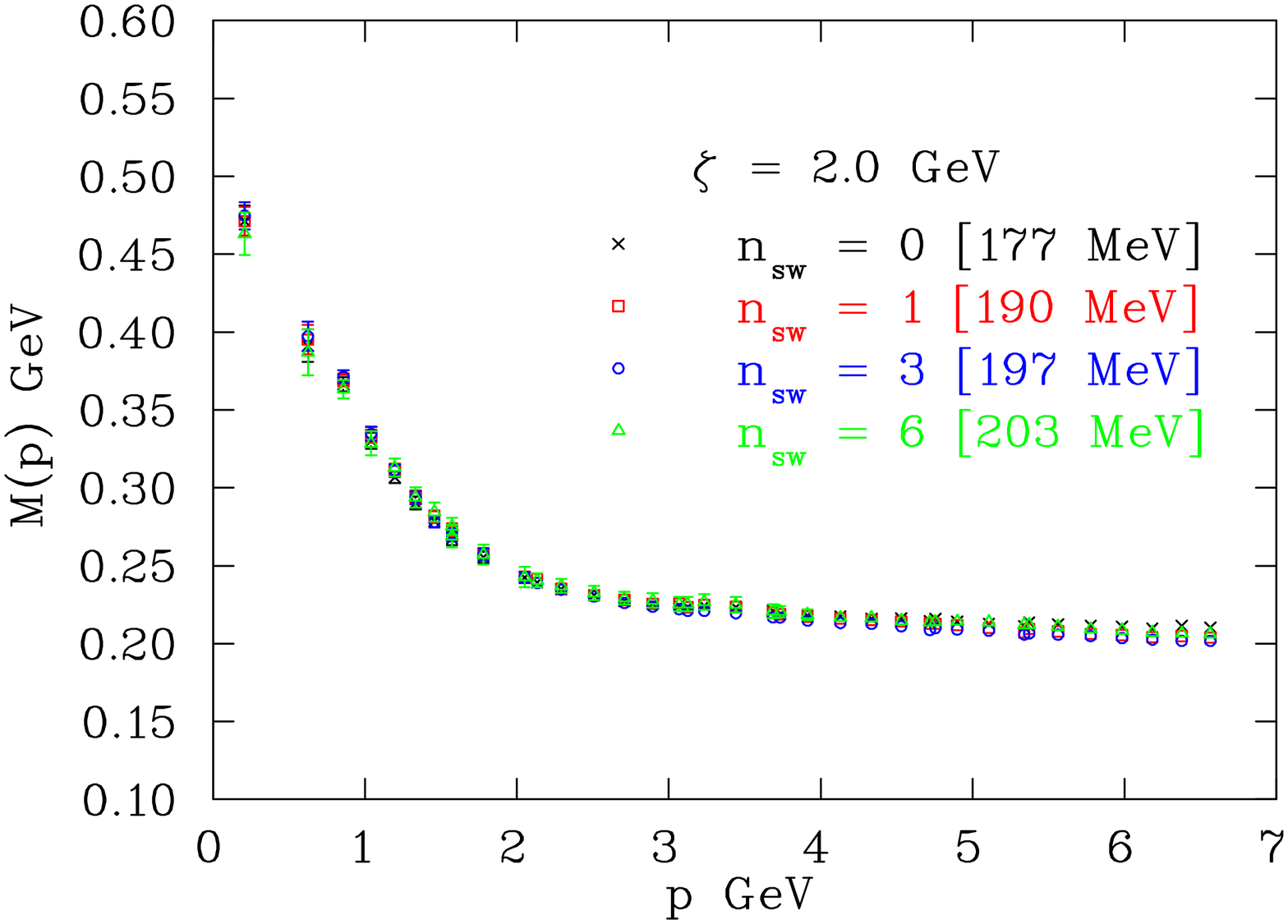} & 
      \includegraphics[angle=90,width=0.36\textwidth]{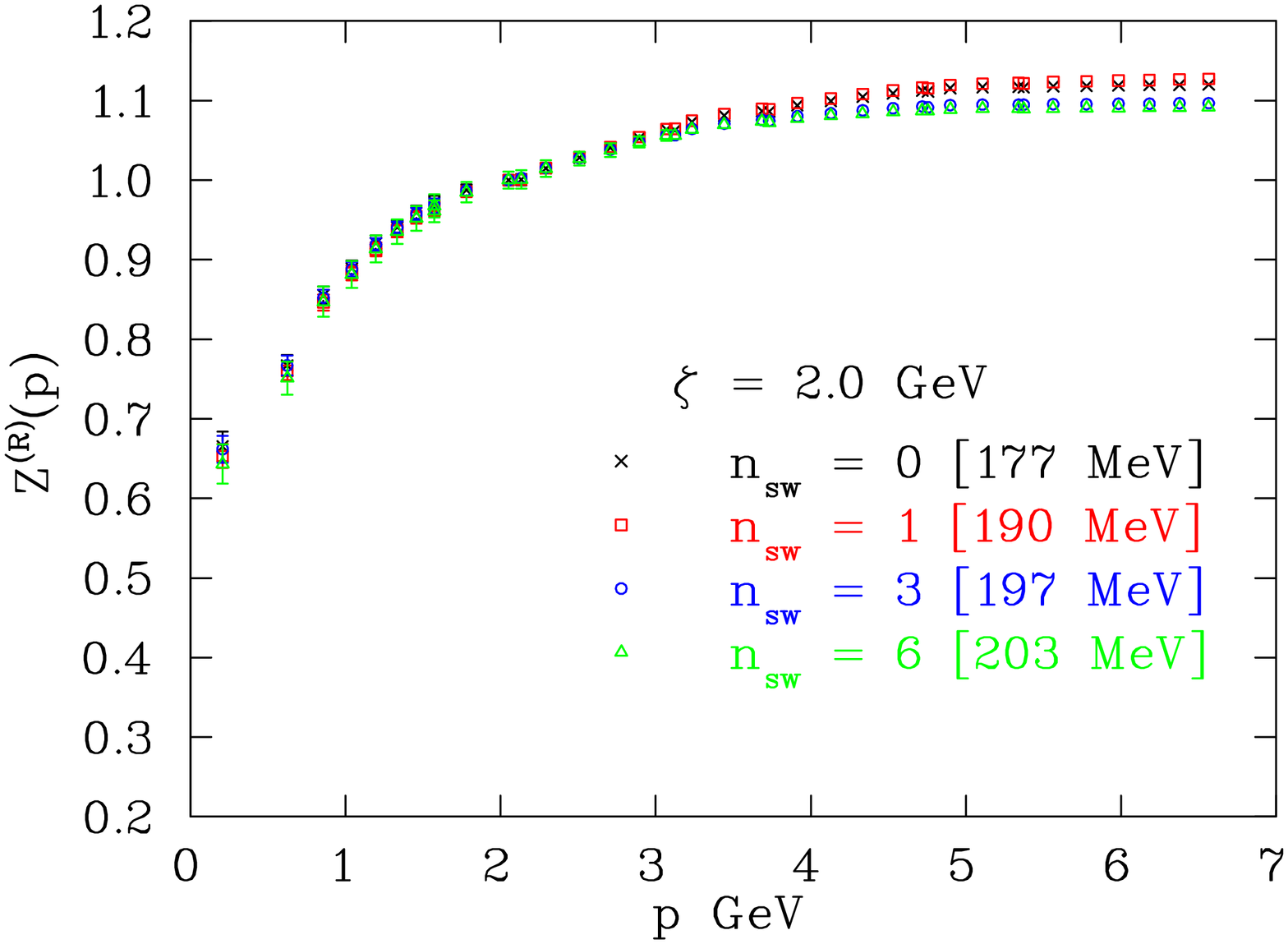} \\
      \includegraphics[angle=90,width=0.36\textwidth]{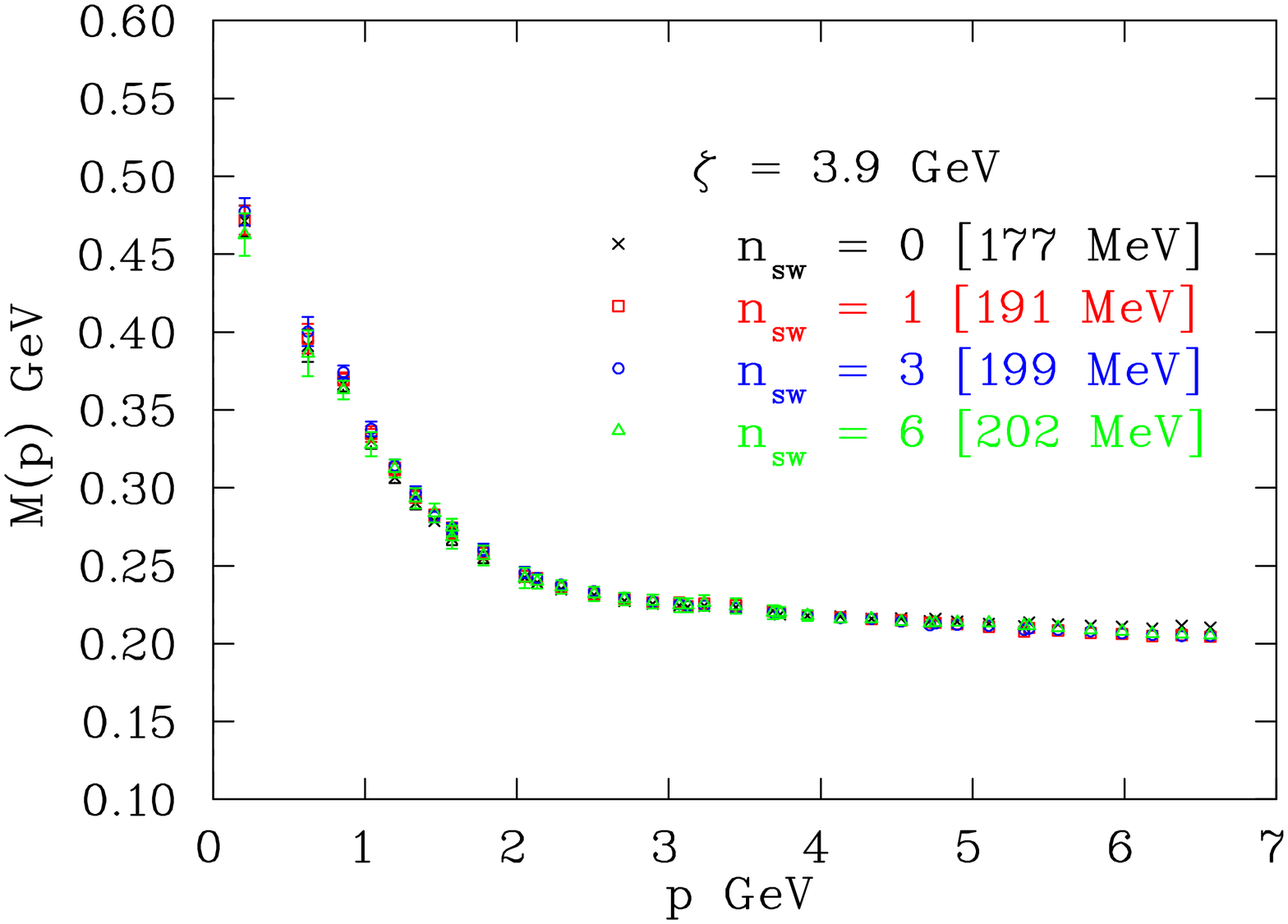} & 
      \includegraphics[angle=90,width=0.36\textwidth]{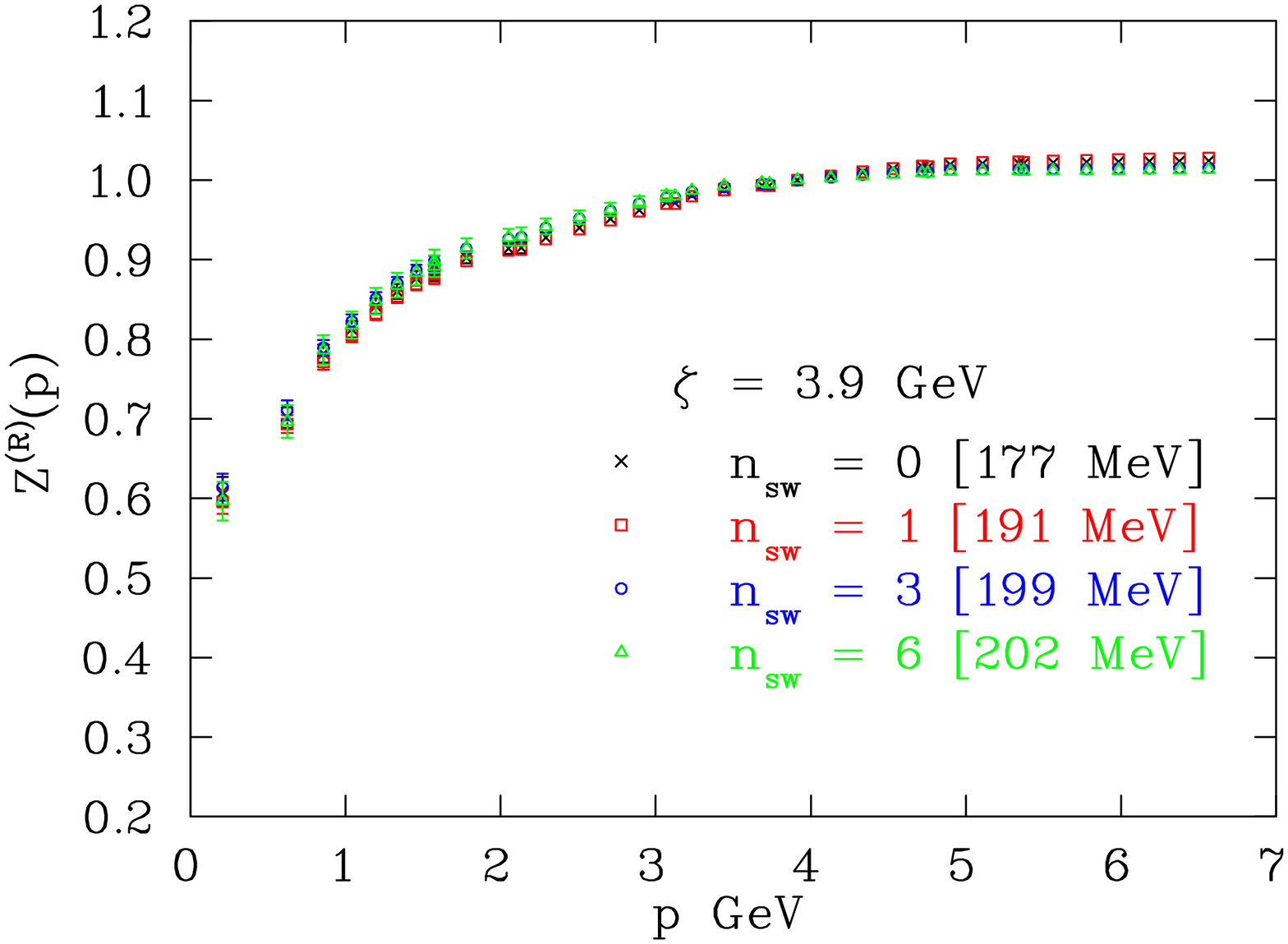} \\
      \includegraphics[angle=90,width=0.36\textwidth]{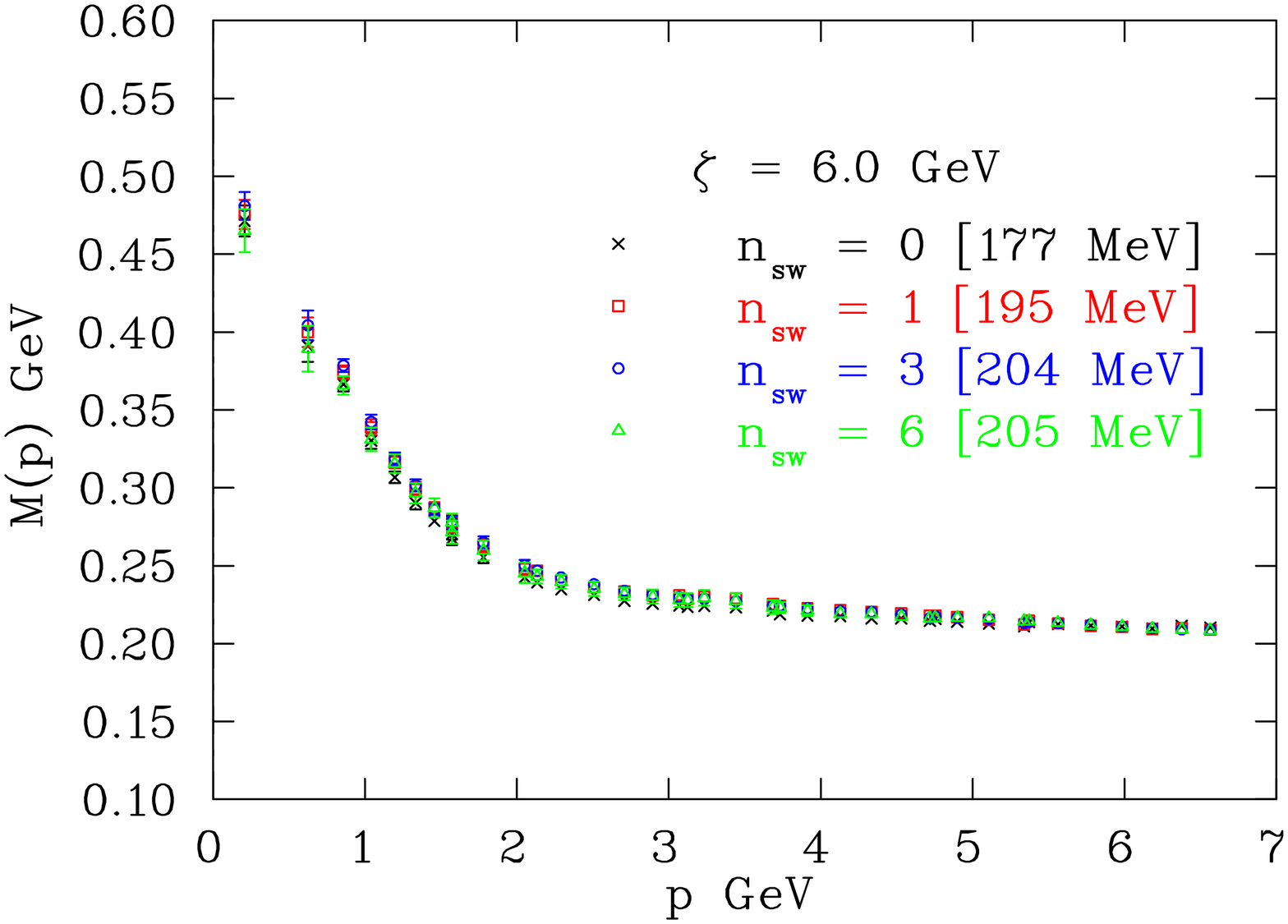} & 
      \includegraphics[angle=90,width=0.36\textwidth]{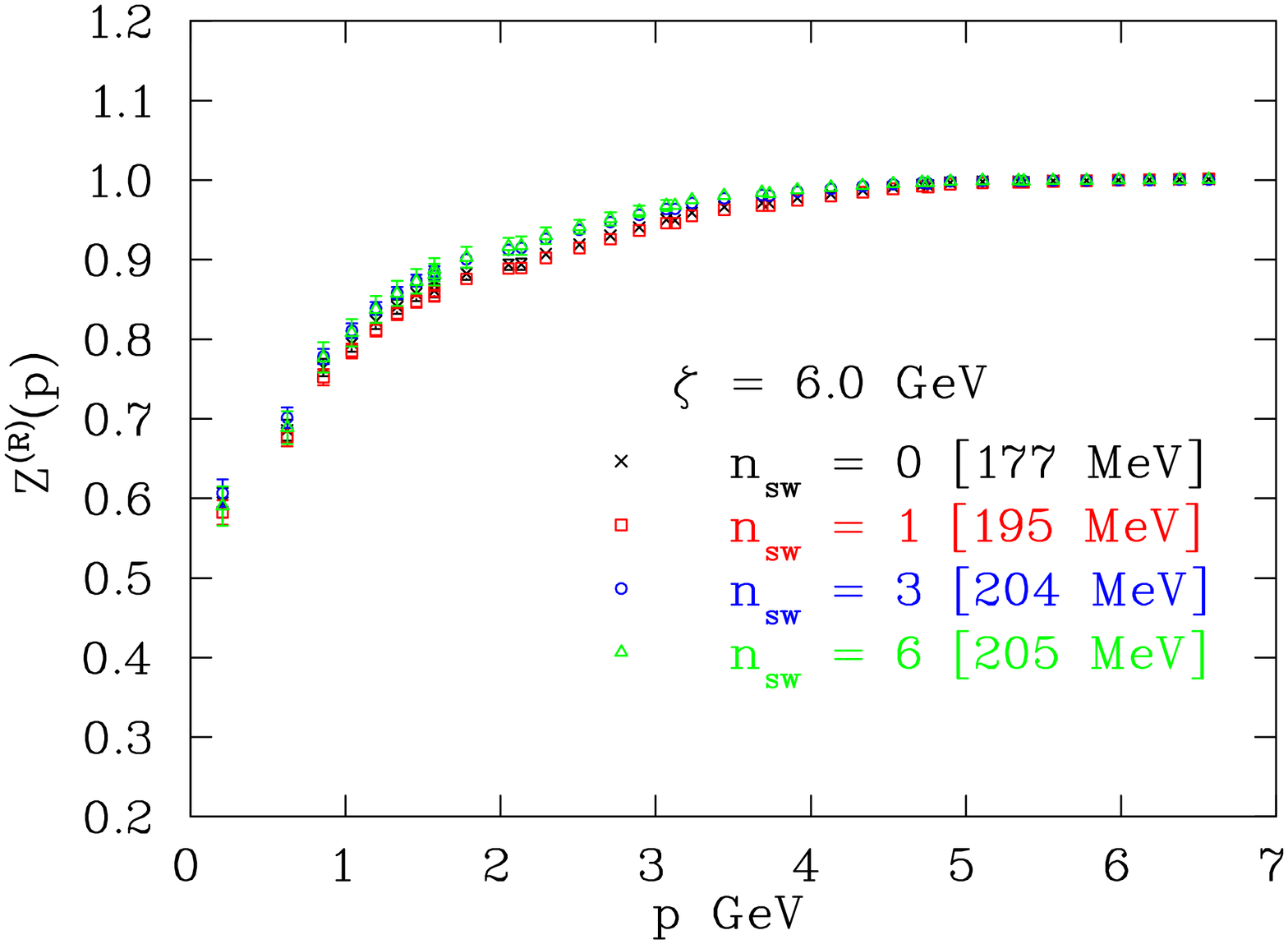}
    \end{tabular}
  \end{center}
\caption{The interpolated mass and renormalization functions for the
  moderate bare quark mass, $m^0 = 177$~MeV, with the three choices of
  $\zeta$. The effective bare quark masses are given in square
  brackets.  Apart from a splitting of the renormalization function
  for $\zeta = 2.0$~GeV, the smearing has little effect on the quark
  propagator.}
\label{moderate}
\end{figure}

\begin{figure}
  \begin{center}
    \begin{tabular}{c@{\hspace{0.05\textwidth}}c}
      \includegraphics[angle=90,width=0.36\textwidth]{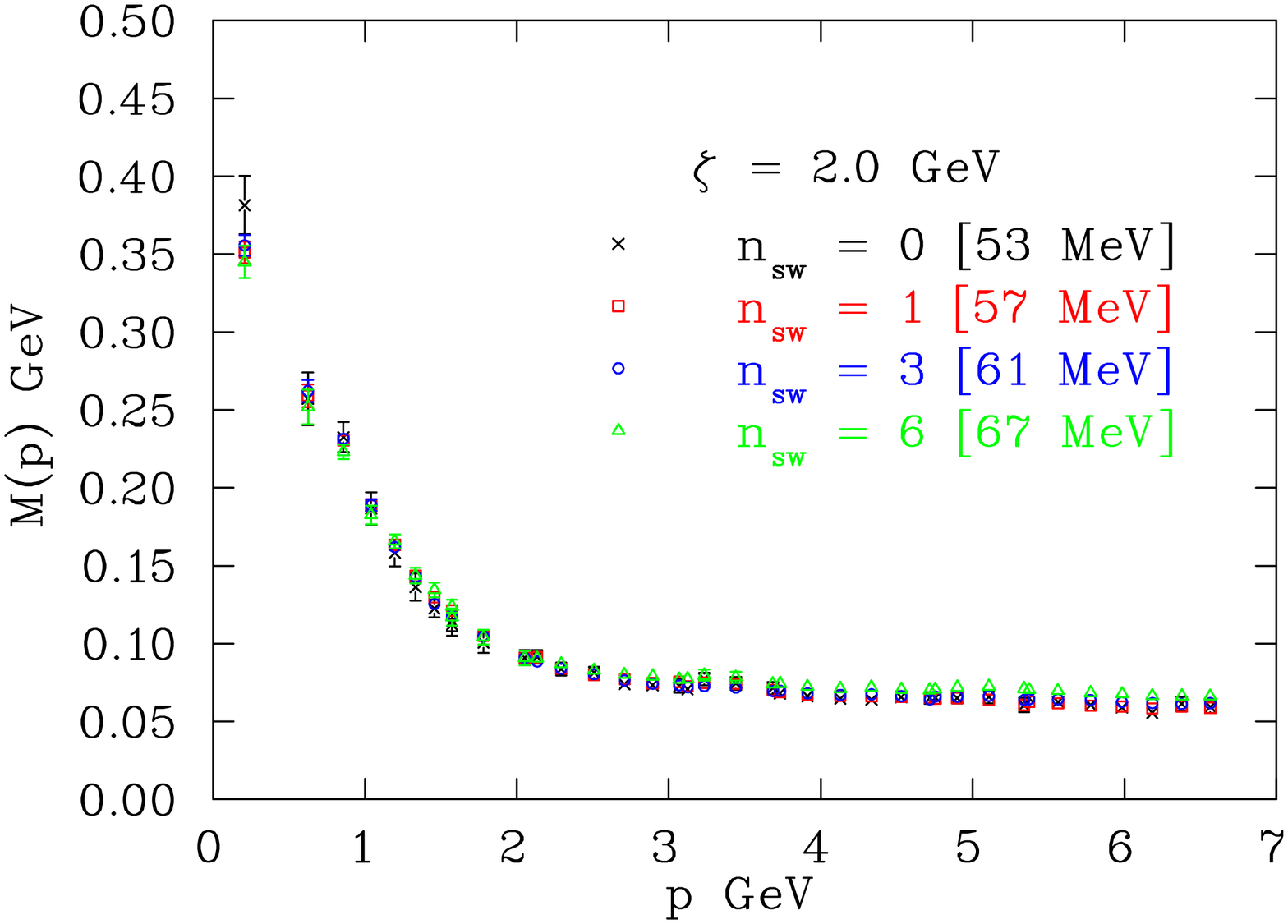} & 
      \includegraphics[angle=90,width=0.36\textwidth]{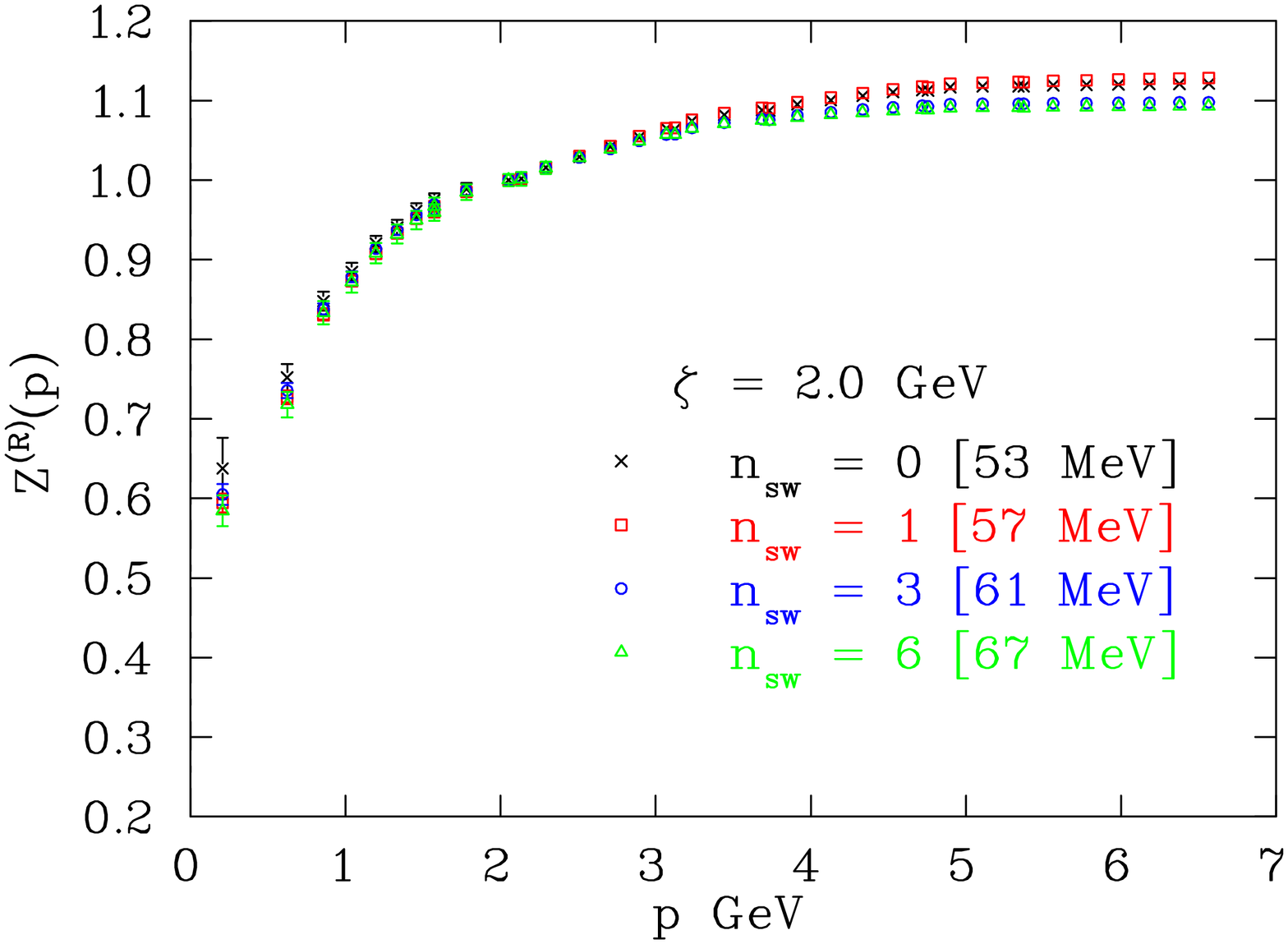} \\
      \includegraphics[angle=90,width=0.36\textwidth]{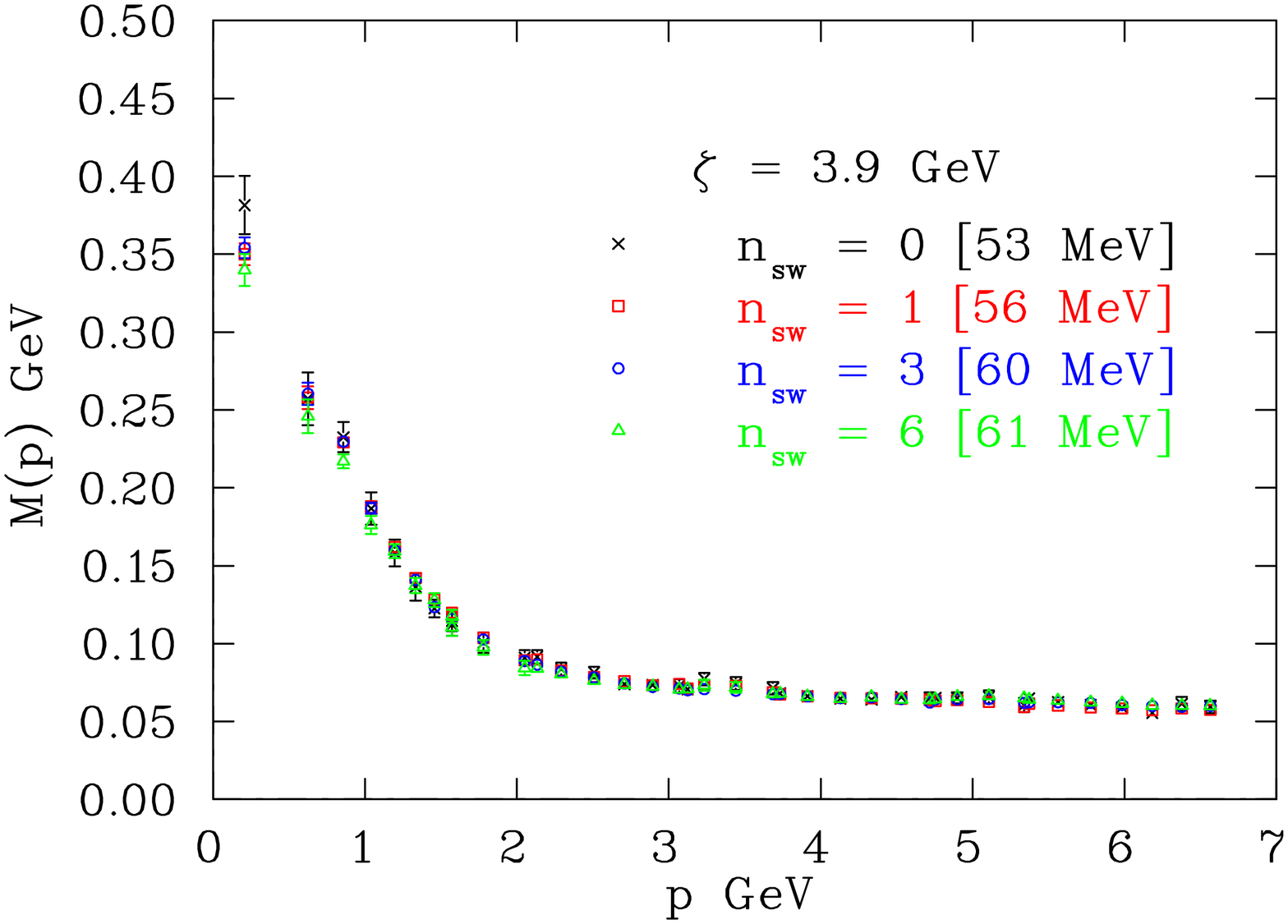} & 
      \includegraphics[angle=90,width=0.36\textwidth]{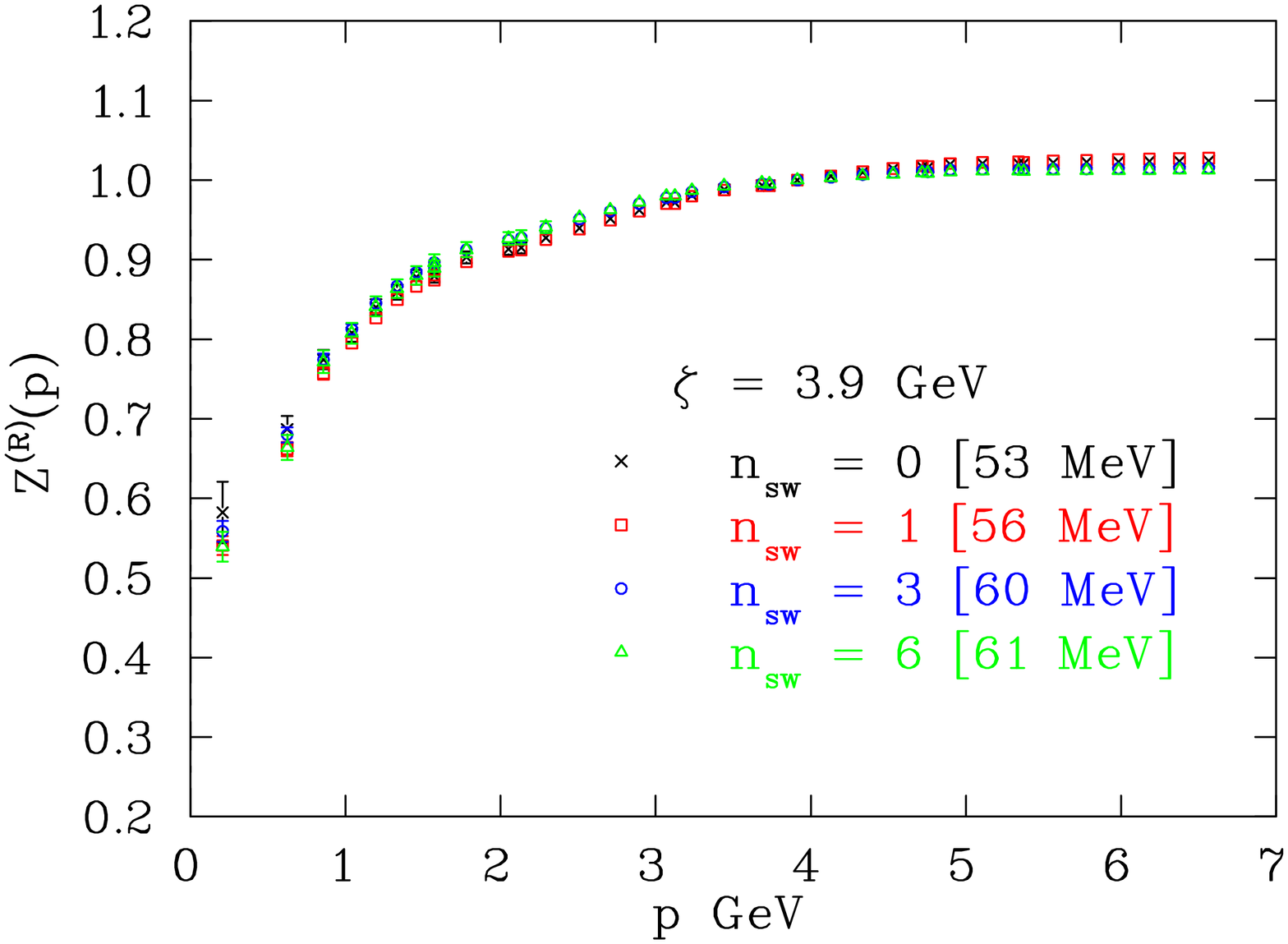} \\
      \includegraphics[angle=90,width=0.36\textwidth]{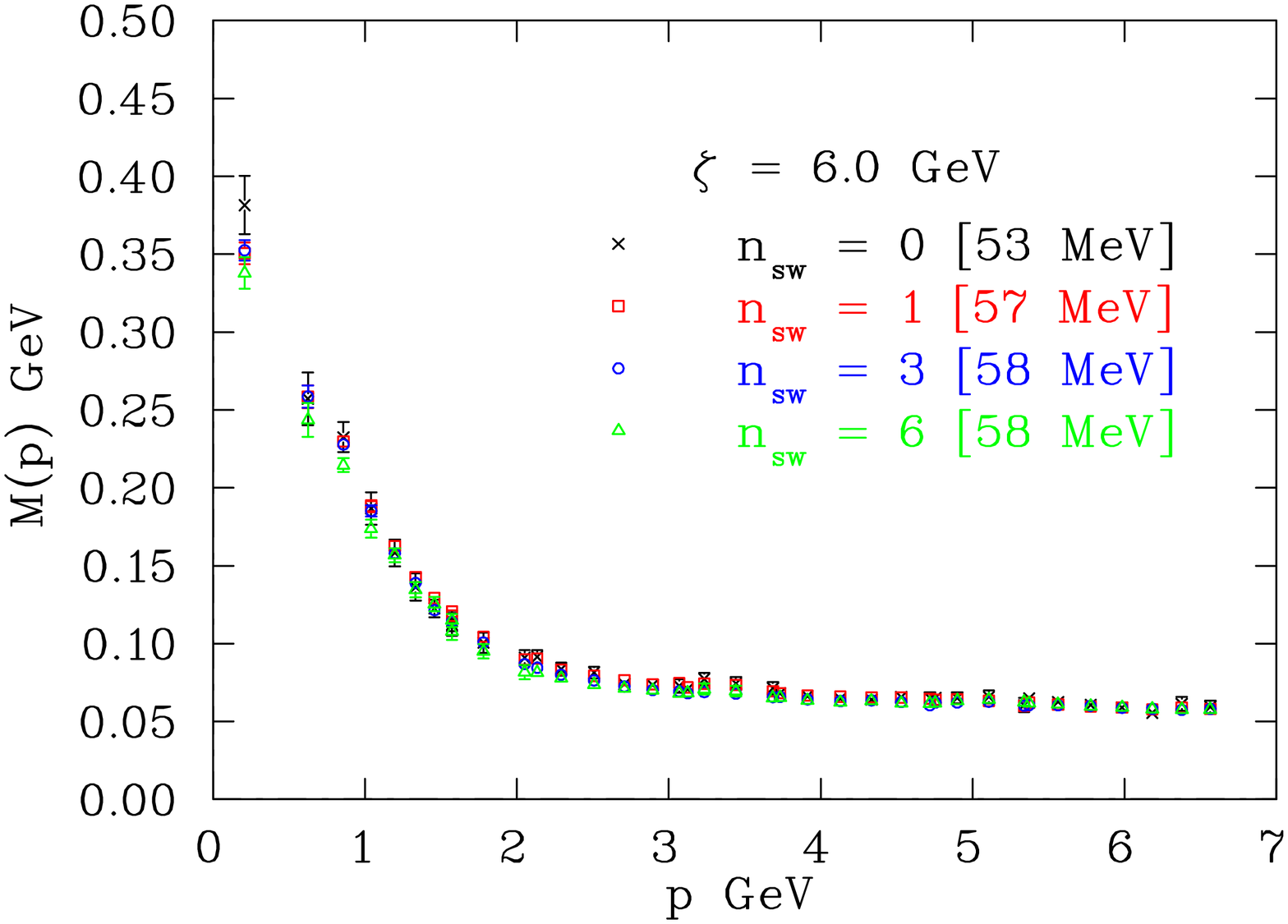} & 
      \includegraphics[angle=90,width=0.36\textwidth]{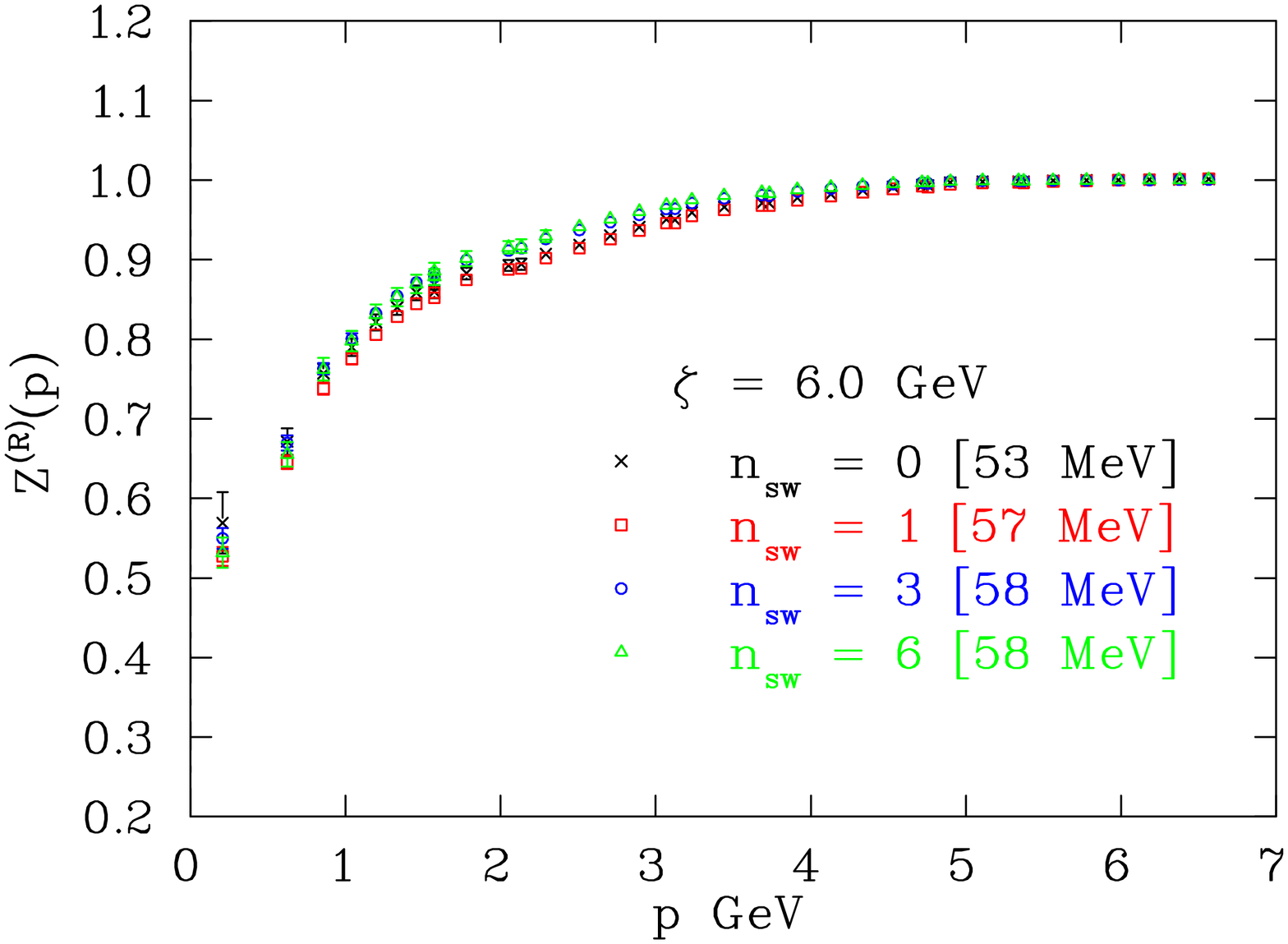}
    \end{tabular}
  \end{center}
\caption{The interpolated mass and renormalization functions for the
  small bare quark mass, $m^0 = 53$~MeV, with three choices of
  $\zeta$.  As with the moderate quark mass, the smearing has little
  effect on the propagator. However when the quark masses are matched
  in the UV at $\zeta = 6.0$~GeV there is a significant suppression of
  dynamical mass generation for $n_{sw} = 6$.}
\label{light}
\end{figure}

\begin{figure}
  \begin{center}
    \begin{tabular}{c@{\hspace{0.05\textwidth}}c}
      \includegraphics[angle=90,width=0.36\textwidth]{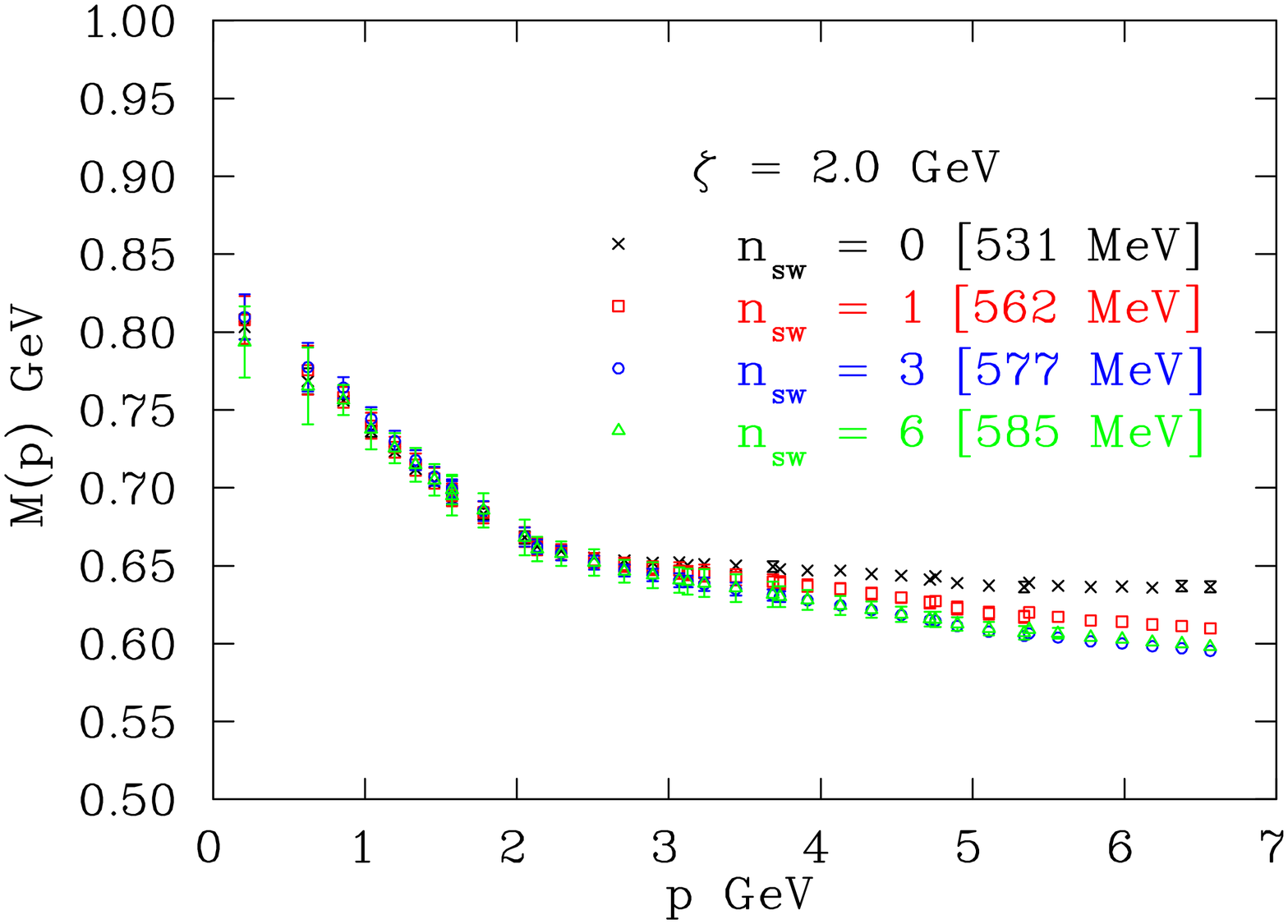} & 
      \includegraphics[angle=90,width=0.36\textwidth]{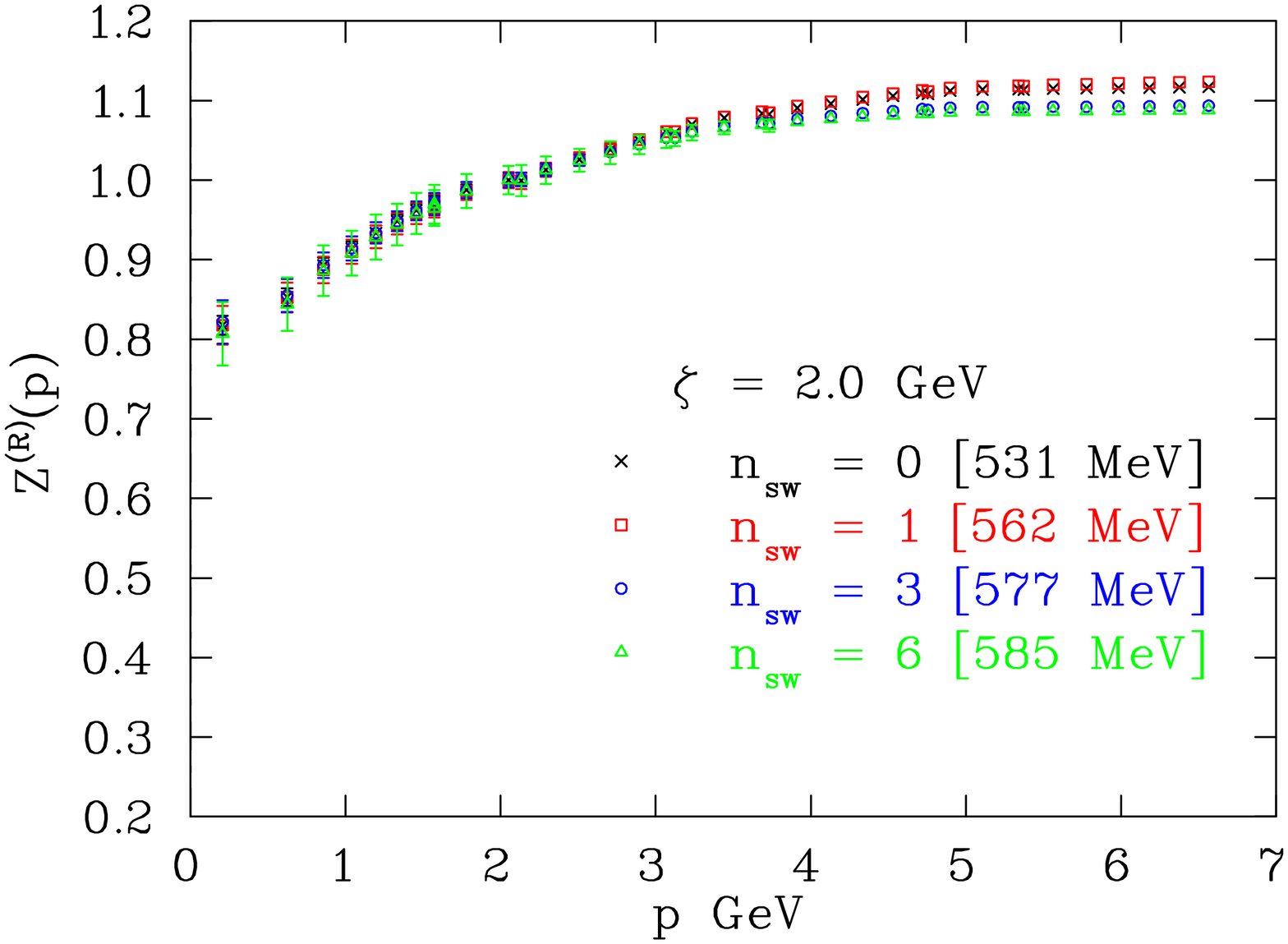} \\
      \includegraphics[angle=90,width=0.36\textwidth]{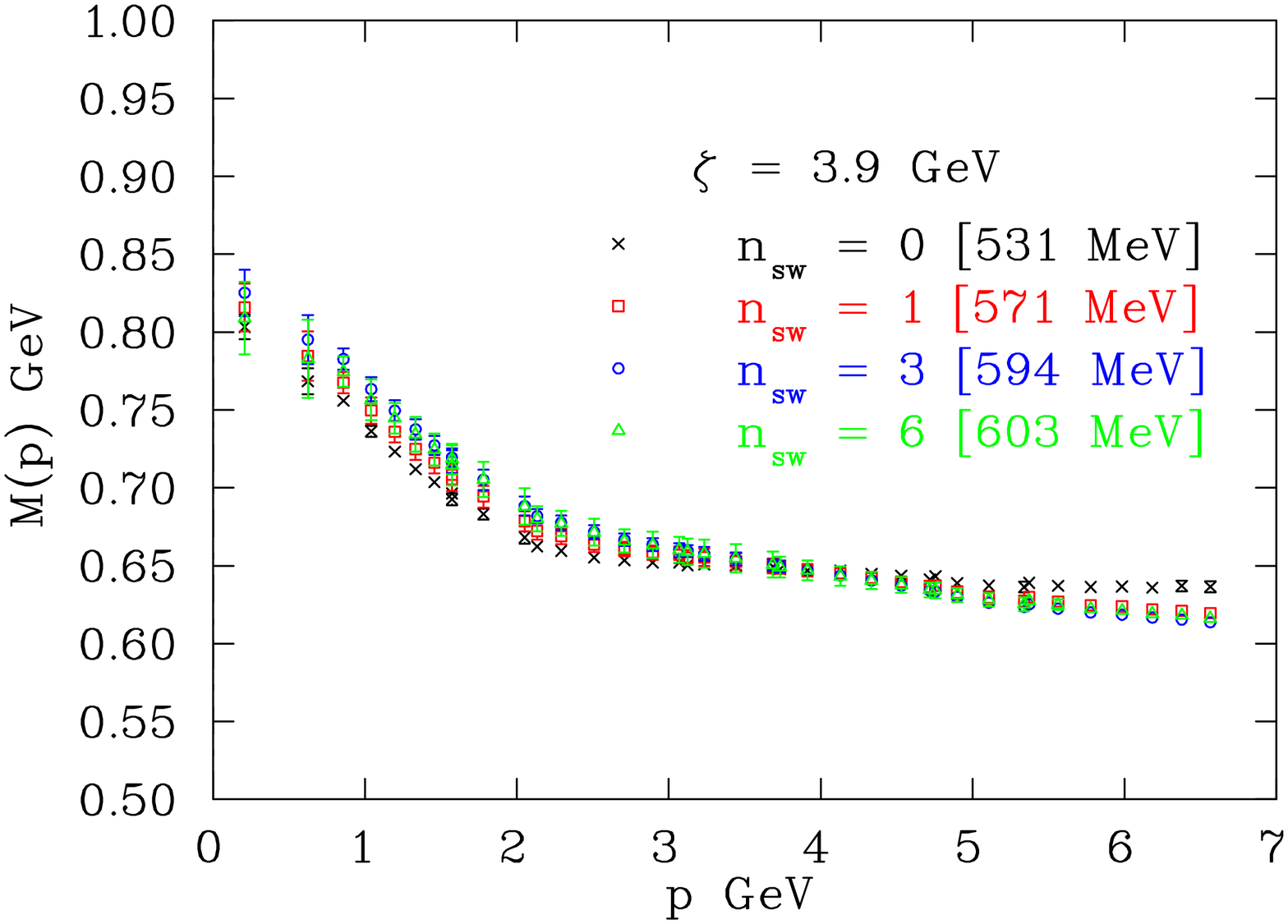} & 
      \includegraphics[angle=90,width=0.36\textwidth]{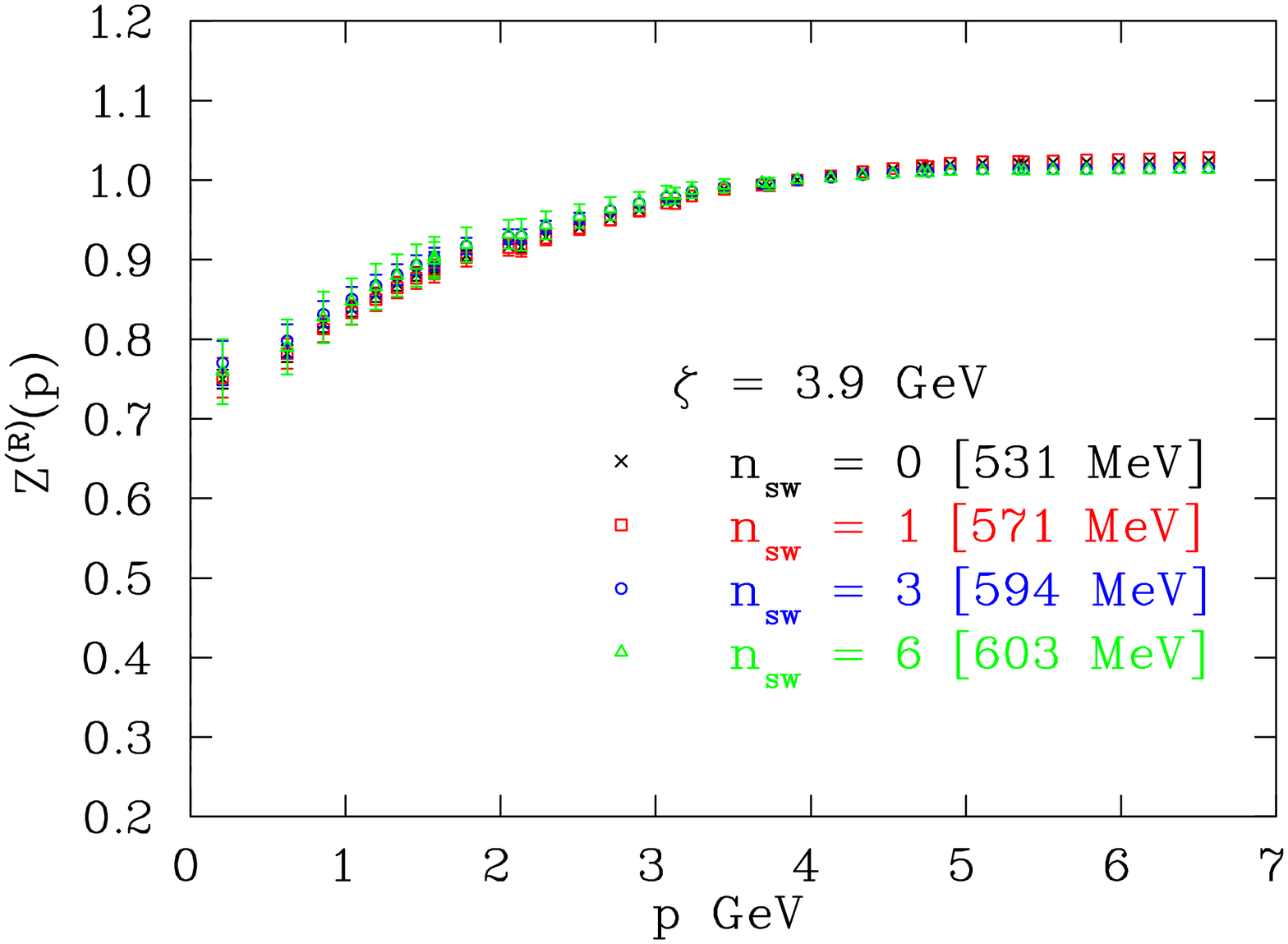} \\
      \includegraphics[angle=90,width=0.36\textwidth]{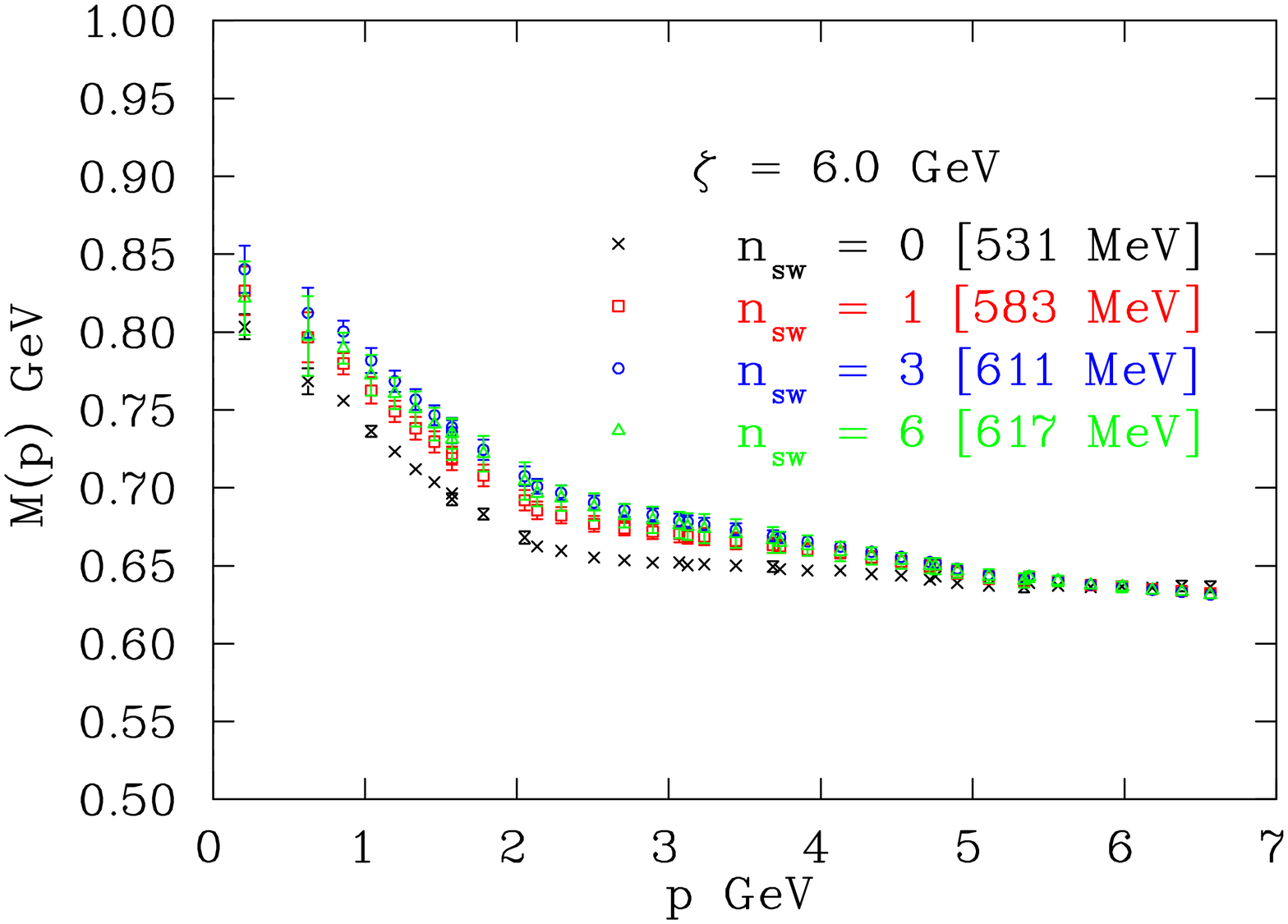} & 
      \includegraphics[angle=90,width=0.36\textwidth]{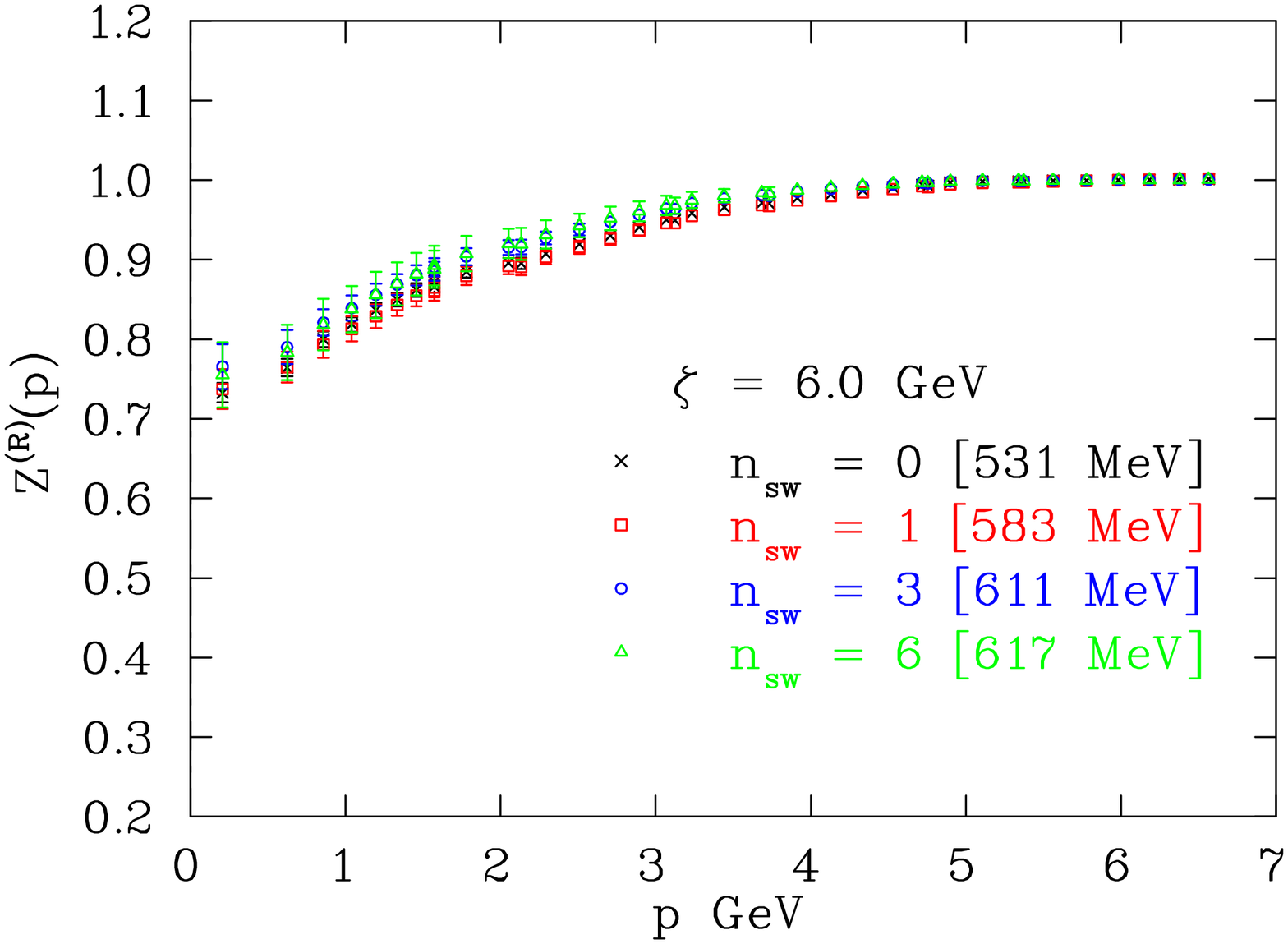}
    \end{tabular}
  \end{center}
\caption{The interpolated mass and renormalization functions for the
  heavy bare quark mass, $m^0 = 531$~MeV, for the three choices of
  $\zeta$. The effective bare quark masses are given in square
  brackets. For this choice of $m^0$ the smearing has a strong impact
  on the quark propagator.}
\label{heavy}
\end{figure}

We begin with an analysis of the moderate bare quark mass, $m_0 =
177$~MeV.  The interpolated mass and renormalization functions, for
three choices of reference momentum, are shown in Fig.~\ref{moderate}.
For this choice of bare quark mass the smearing algorithm appears to
have little effect on the quark propagator.  The most obvious
dependence is in the UV region of the renormalization function for
$\zeta = 2.0$~GeV, where the smearing introduces a splitting of the
curves.

Next we consider the lighter quark mass, $m_0 = 53$~MeV.  The mass and
renormalization functions are given in Fig.~\ref{light}.  As in the
case of the moderate quark mass there appears to be little effect on
the quark propagator.  However when the quark masses are matched in
the UV at $\zeta = 6.0$~GeV there is a significant suppression of
dynamical mass generation for $n_{sw} = 6$.

Finally, we consider the heavy quark mass, $m_0 = 531$~MeV, and
provide the relevant functions in Fig.~\ref{heavy}.  The impact of
smearing is most apparent for this heavy mass.  The most interesting
effect is in the mass function with the choice of $\zeta = 2.0$~GeV,
matching the infrared physics.  Here the mass function decreases
dramatically as progressively more smearing is applied.  We see that
smearing the gauge field causes the asymptotic value of the mass
function to approach the input bare quark mass, clearly illustrating
the short-distance void created by 6 sweeps of stout-link smearing.

\section{Conclusion}

Smearing all links in the gauge field only has a strong effect on the
quark propagator for heavy quarks.  Here suppression of the short
distance physics through smearing spoils the physics of the theory
above about 2 to 3~GeV.  Renormalization of the bare quark mass is not
enough to restore the flattening of the bare quark mass at large
momenta.  We also saw that the asymptotic value of the mass function
approaches the input bare quark mass.

For all values of the quark mass considered, the effect of one sweep
of smearing on the renormalization function is negligible.  The small
discrepancies, which are of the order of 2\%, are insensitive to the
value of the quark mass.  We also note a significant reduction in the
statistical error, even after a single sweep of smearing.

At the lightest quark mass, the anticipated suppression of dynamical
mass generation is apparent.  In the case of six smearing sweeps, we
find an order $2\sigma$ reduction in the infrared dynamical mass
generation when the quark masses are matched.  In this case, we also
see that for $p < 2$ to $3$~GeV, the mass function for six sweeps of
smearing lies systematically low.  This effect is clearly illustrated
in the lower left panel of Fig.~\ref{light}.

\acknowledgments

We thank both eResearch SA and the NCI National Facility for generous
grants of supercomputer time which have enabled this project.  This
work is supported by the Australian Research Council. J. B. Zhang is
partly supported by Chinese NSFC-Grant No.~10675101 and 10835002.  POB
is supported by the Marsden Fund administered by the Royal Society of
New Zealand.

\end{document}